# TTSS Packet Classification Algorithm to enhance Multimedia Applications in Network Processor based Router

R.Avudaiammal ,
Research Scholar,
Anna University
avudai_r@yahoo.com ,

R.SivaSubramanian ,
Research Scholar,
Vinayaka Mission's University
usharss@yahoo.com,

R.Pandian      and P. Seethalakshmi
Research Scholar,
Anna University         Anna University
rpirtt@yahoo.co.in,    auropansee@yahoo.co.in

***Abstract-*** *The objective of this paper is to implement the Trie based Tuple Space Search(TTSS) packet classification algorithm for Network Processor(NP) based router to enhance multimedia applications. The performance is evaluated using Intel IXP2400 NP Simulator. The results demonstrate that, TTSS has better performance than Tuple Space Search algorithm and is well suited to achieve high speed packet classification to support multimedia applications.*

**Key words -** Multimedia, QoS, Packet Classification, TTSS, Network Processor, IXP 2400.

## I. INTRODUCTION

Multimedia traffic has become high due to the growth of multimedia applications such as video streaming, video conferencing, IP telephony and e-learning. These Multimedia applications require better Quality of service which can be provided by properly classifying and scheduling the packets, shaping the Traffic, managing the bandwidth, and reserving the resources. All these techniques require the fundamental task Packet classification at the router, to categorize the packets into flows. The router classifies the packets into flows, by comparing the L2 / L3 / L4 fields in the header of the incoming packet with the filters that are stored in the router table where a Filter (F) is a set of rules defining the packets attributes for the routing purpose. The classifier splits the incoming packet header in the same way as filter field and classifies the packets into flows. If the algorithm uses more than one field of the packet header, it is called multidimensional Packet Classification. 5-tuple header field consisting of IP source and destination address, protocol type source port and destination port are the common fields under literature even though higher number of fields is possible for applications requiring QoS [1]. Packet Classification (PC) involves complex tasks with various matching conditions, such as Longest Prefix Matching (LPM)( for source and destination address prefix), Exact Matching (EM)(for protocol), Range Matching (RM)(for port) and choosing the best rule. As the packet classification has to be done at wire speed in the router path it becomes a bottle neck at high speed interfaces.

Present Routers are mainly based on Application Specific Integrated Circuits (ASICs) that are custom made and are not flexible enough to support diversified services. General Purpose Processors(GPP) offer flexibility in supporting new features by simply upgrading the software, but have difficulties in supporting higher bandwidth[1]. Network Processors have emerged to provide both the performance of ASICs and the programmability of GPPs. Powerful Network Processors have been introduced by many companies that can be placed in routers to execute various network related tasks at packet level that supports QoS functionalities. The design and development of routers using NP has gained significance due to its high performance. Hence, computationally intensive task of Packet Classification can be implemented in line speed using Embedded Network Processors to enhance QoS.

The work presented in this paper is based on the fully programmable Intel ® IXP 2400 processor. It has a set of hierarchically distributed memory devices, a set of optimized instructions to carry out packet level parallel processing operations. Network Processors allow multitasking and multithreaded programming. Each processing element (PE) has multiple hardware thread contexts that enable thread context switches that have zero or minimal overhead [2]. It can examine and forward packets independently without using the host processor, bus, or memory. This paper proposes Trie based Tuple Space Search(TTSS) packet classification algorithm using IXP 2400 to increase the throughput and packet classification rate that supports multimedia applications.

The remainder of this paper is organized as follows: section II presents the brief description of the Intel IXP2400 architecture and Section III presents the background concept of packet classification .The performance analysis is presented in section IV and conclusion in Section V.

## II. INTEL IXP2400 NETWORK PROCESSOR

The Intel® IXP2400 is a member of the Intel's second generation network processor family. Network processors are designed to perform a wide range of functionalities such as multi-service switches, routers, broadband access devices and wireless infrastructure systems. It implements a high-performance parallel processing architecture on a single chip that is suitable for processing complex algorithms, detailed packet inspection, traffic management and forwarding at the wire speed.

The architecture of an integrated network processor IXP2400NP shown in Figure 1 has a single 32-bit XScale core processor, eight 32-bit MicroEngines (MEs) organized as two clusters, standard memory interfaces and high speed bus interfaces. Each microengine has 256 general purpose registers, that are equally shared between eight threads. Microengines exchange information through an on-chip scratchpad memory or via 128 special purpose next neighbor registers. Data transferring across the MEs and locations external to the ME, (for eg DRAMs, SRAMs etc.) are done by the available 512 Transfer Registers The Xscale core is responsible for initializing and managing the chip, handling control and management functions and loading the ME instructions . Each ME has eight hardware-assisted threads of execution with no context switch overhead. All threads in a particular ME execute code from the instruction stored on that ME, whose size is 4K. The SRAMs and DRAM are off chips that are shared by all processors. In general, SRAM is used for storing the table information such as routing table and DRAM is used for packet buffering. Also the IXP2400 chip has a pair of buffers (BUF), Receive BUF (RBUF) and Transmit BUF (TBUF) that are used to send / receive packets to / from the network ports with each of size 8 Kbytes. The data in RBUF and TBUF is divided into sub blocks referred to as elements.

The other noteworthy features of the IXP2400 architecture include a hash unit, a scratchpad memory, a Control Status Register (CSR) Access Proxy, a Peripheral Computer Interface (PCI) controller , a Media Switch Fabric (MSF) Interface and DRAM and SRAM controllers.

The packets are injected into the Network Processor from the network through the Media Switch Fabric Interface. Then the packet data and its meta data are kept in DRAM and SRAM respectively. The packets are then forwarded to the MicroEngines for processing .Finally the processed packets are driven into the network by Media Switch Fabric Interface (MSF) at output port.

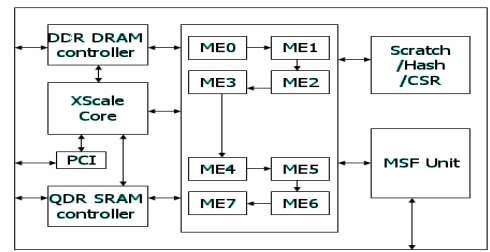

Figure 1. Architecture of IXP2400

## III. RELATED WORK

Packet classification entails searching a table of filters which binds a packet to a flow or set of flow and returning the flow id (action applied to the packet). Survey on packet classification algorithms [4] and [5] shows that the problem is inherently hard. Longest Prefix Matching for routing lookups [7] is a special case of packet classification which has the advantage of making algorithm faster and less costly to implement.

Linear search is the simplest method of packet classification that uses linked list to perform searching of each packet through a set of filters sequentially. The spatial and temporal complexity is O(N) where 'N' is the number of rules in the rule set . It exhibits poor scaling. By exploiting the redundancy of the classifiers in real networks, multi field classification can be performed with lesser number of memory access and storage. Tuple space is a heuristic approach for packet classification [6] and it attempts to quickly narrow down the scope of a multiple field search by partitioning the filter set by "tuples". A tuple defines the number of specified bits in each field of the filter. The set of rules mapped to the same tuple can be stored in a Hash table and the set of tuples is called a "tuple space". As the number of distinct tuples is much lesser than the number of filters in filter set, even a simple linear search of the tuple space provides significant speed up than the linear search over the filters. If 'w' is the length of the IP prefix and 'k' is the number of classified fields then its search time is $O(w^k)$ and its memory capacity is O(N).

Some of the trie-based algorithms follow the hierarchical approach of the packet classification which recursively performs search in each field. Using the destination prefix field, all matching candidate rules are identified, and for those candidates, rules are further filtered using the source prefix field, protocol and port number. In [18-21], trie based classification algorithm suitable for high speed two dimensional packet classification have been described. A one-dimensional 1-bit trie is a binary tree like structure, in which each node has two element fields, le(the left element) and re (the right element), and each element field has the components child and data. Branching is done based on the bits in the search key. A left-element child branch is followed at a node at level i (the root is at level 0) if the $i^{th}$ bit of the search key is 0; otherwise, a right element child branch is followed. In One-Dimensional Multibit Tries the term 'stride' of a node is defined as the number of bits used at that node to determine which branch is to be taken. A node with stride 's' has $2^s$ element fields and requires $2^s$ memory units (one memory unit being large enough to accommodate an element field). Note that the stride of every node in a 1-bit trie is 1.The enhancement of one-dimensional 1-bit trie is two-dimensional 1-bit trie (2D1BT) .It has two levels in which the data field of each element of the top-level one-dimensional 1-bit trie is a pointer to a lower-level 1-bit tries. Two-dimensional multibit tries (2DMTs) are a natural extension of 2D1BTs where the stride of the destination and source tries is more than 1. As in the case of one-dimensional multibit tries, prefix expansion is used to accommodate prefixes whose length lies between the [s, e] values of a node on its search path.

Trie based Tuple Space Search (TTSS) is an extension of 2DMT .Each node has multiple element fields. In TTSS, the term stride is equivalent to prefix . where each element is a pointer of one hash table. Left most element of the node has the hash table of longest prefix length. Since all prefixes in a group have the same length, the length of prefix bit string can be used as a key to decide which branch is to be taken. Its memory capacity is O(N) since each rule is stored in exactly one Hash table.

Directions in designing the classification algorithms is introduced [9] using network processors and other studies [10-17] have focused on implementing the networking services using programmable network processors.

Though there are several algorithms specialized for the case of rules on two fields (e.g. Source and destination IP address only), it is necessary to select a packet classification algorithm that requires both low memory space and low access overhead. In summary, for the general classification problem on three or more fields, TTSS could be effective . Hence, in this paper Trie based Tuple Space search algorithm with two versions one with hash table of lengthy prefix as the left most element TTSSV1and the other with hash table of lengthy prefix as the right most element TTSSV2 on the IXP 2400 platform have been implemented and evaluated .

## IV. IMPLEMENTATION

The framework for the implementation of the algorithm is shown in Figure 2.

Packet Receive Microengine has been interfaced with the Media Switch Fabric Interface (MSF). The packets are injected through Media switch Fabric Interface and Receive Buffer(RBUF) reassemble incoming m-packets(packets with length 64 bytes). For each packet, the packet data is written to DRAM, the packet meta-data (offset, size) is written to SRAM. The Receive Process is executed by microengine (ME0) using all the eight threads available in that microengine (ME0). The packet sequencing is maintained by executing the threads in strict order. The packet buffer information is written to a scratch ring for use by the packet processing stage Communication between pipeline stages is implemented through controlled access to shared ring buffers.

Classifier microblock is executed by microengine (ME1) that removes the layer-3 header from the packet by updating the offset and size fields in the packet meta descriptor. The packets are classified into different traffic flows based on the IP header and is enqueued in the respective queue.

Then the Packet Transmit microblock segments a packet into m-packets and moves them into TBUFs for transmitting the packets over the media interface through different ports. The packet Transmitter microblock monitors the MSF to see if the TBUF threshold for specific ports has been exceeded. If so it stops transmitting on that port and any requests to transmit packets on that port are queued up in local memory. The Packet Transmitter microblock periodically updates the classifier with information about how many packets have been transmitted. In this paper, microblocks are implemented using microcode.

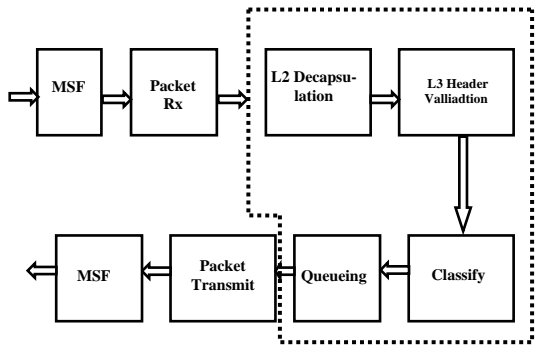

Figure 2. Software framework Design

*Implementation Environment*

IXA 3.51 SDK is a cycle based simulator [15] in which IXP2400 is set to run under the following conditions: PLL output frequency: 1200 MHz, ME frequency: 600 MHz, Xscale Frequency: 600 MHz. SRAM frequency: 200 MHz, two channels, 64 MB per channel, DRAM Frequency :150 MHz, 64 MB. The configuration for the device type is x32 MPHY4 with bus mode 1x32 which controls how packets are sent and received from the simulator. A device with 4 ports, each with a data rate of 1000 Mbps and receive and transmit buffer sizes of 65536 and 16384 respectively are chosen for this application. The simulator is configured to send packet streams to ports 0 through 3 of the device. Implementation assigns individual blocks from the fast path pipeline to separate microengines on the IXP2400 NPs.

Multimedia stream is artificially simulated with UDP flows since voice and video applications use UDP rather than TCP. Traffic generated in this work consists of RTP/UDP packets, UDP packets (with large TTL and with less TTL) and TCP packets. All these packets are uniformly distributed in the traffic. Traffic is generated with a constant rate of 1000Mb/s and inter-packet gap is set to 96 ns.

V. PERFORMANCE EVALUATION

In this paper, 5-tuple header field consisting of IP source and destination address, protocol type, TTL and ToS is considered. These five fields are common fields under literature even though higher number of fields is possible [10]. The classifier splits the incoming packet header in the same way as filter field and classifies the packets into four flows $F_1$ (RTP), $F_2$ (delay sensitive UDP packets with less TTL and ToS), $F_3$ (delay sensitive UDP packets with large TTL) and finally flow $F_4$ (TCP packets). In general, prefix match is used for IP source and destination addresses and exact match is for protocol flag. Due to smaller number of application specification fields (protocol type), the filter set is searched starting with protocol field to minimize the number of entries to be searched.

The packet classification algorithms such as Tuple Space Search (TSS) and Trie Based Tuple Space Search (TTSS) in two different versions have been implemented using single microengine

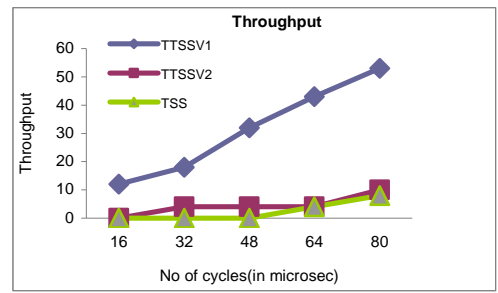

Figure 3. Throughput Vs various algorithms

Figure 3 shows that at the end of 50,000 microengine cycles (80 micro sec), the number of packets classified in TTSSV1 is 81% and 84% more than the TTSSV2 and TSS respectively

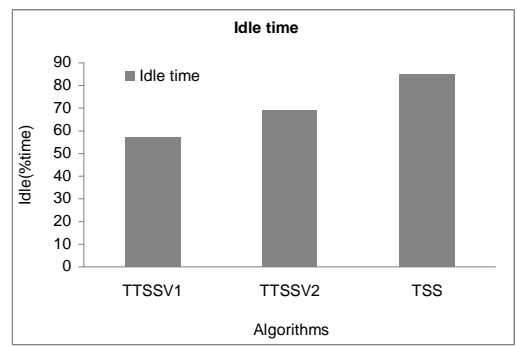

Figure 4. Microengine Idle time

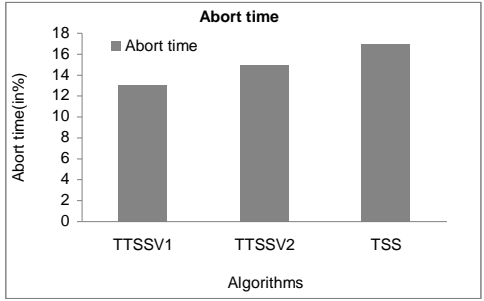

Figure 5. Microengine Abort time

ME is the important resource of NP whose utilization factor can be described by its idle time and Aborted time. Figure 4 shows that reduction in Idle time of Microengine ME0:1 is about 17% and 33% in TTSSV1 classification algorithm when compared to TTSSV2 and TSS respectively. Aborted time of Microengine ME0:1 is the percentage of a total time that was wasted due to instructions that is being aborted. Figure 5 depicts that the Aborted time of the Microengine( ME0:1 ) in TTSSV1 classification algorithm is 13% and 23% lesser than that of TTSSV2 and TSS respectively and implies that Microengine involved in packet classification has been utilized very effectively.

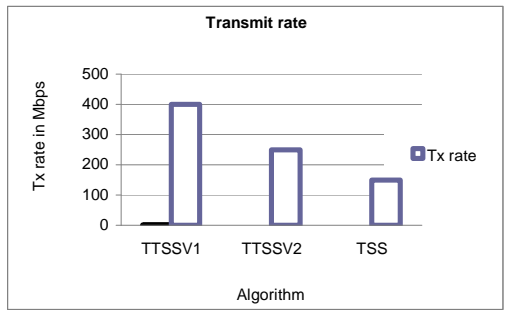

Figure 6. Transmit rate

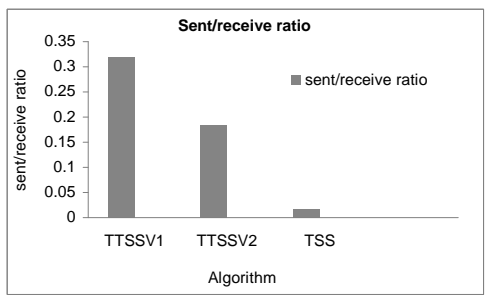

Figure 7. Sent/receive ratio

Figure 6 shows that Transmit rate of the Network processor with fixed Receive rate ,in TTSSV1 algorithm is 38% and 62.5% more when compared with TTSSV2 and TSS respectively. Figure 7 illustrates that packets Sent/Receive ratio of TTSSV1 is 43% and 95% more than TSSV1 and TSS respectively. On summarizing the analysis, Trie based Tuple Space Search performs the high speed packet classification and hence is suitable for multimedia applications.

## VI. CONCLUSION

Packet classification is an important function performed by network devices such as edge router, firewalls and intrusion detection systems. These devices can utilize programmable Network Processors (NP) to implement a classification algorithm and perform this computationally intensive task at line speeds. Trie based Tuple Space Search(TTSS) packet classification has been implemented in this paper. The results show that Throughput is 84% more , Idle time is 33% lesser and packets Sent/Receive ratio is about 95% more than TSS Thus Trie based Tuple Space search (TTSSV1) packet classification in Network processor based router is very effective in terms of Throughput, Microengine Utilization factor for multimedia applications.

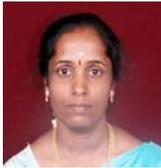

**Mrs. R. Avudaiammal** has received her B.E. degree in Electronics and Communication Engineering from Madurai Kamaraj University, India in 1992 and M.E. degree in Applied Electronics from Bharathiar University , India in 2000. She is an Assistant Professor at St.Joseph's College Of Engg, Chennai ,India. She has 16 years of Teaching experience. She has published books on Microprocessors and on Information coding Techniques. She is currently pursuing her research at Anna University Trichy, India . Her research interest are in Embedded systems, Multimedia Networks and Network Processor ..

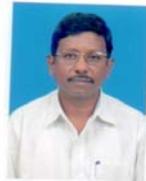

**Mr. R. Pandian** has obtained his B.E. degree in Electrical and Electronics Engineering from Annamalai University, India. in 1981 and M.E. degree in Power Systems Engineering from Madurai Kamaraj University, India in 1983. He has 25 years of Teaching experience and is currently pursuing his research at Anna University Chennai, India. His areas of interest are Computer Architecture, Microprocessors, Microcontrollers and Network Processors.

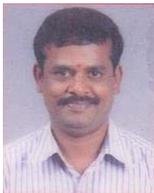

**Mr. R. Sivasubramanian** has obtained his B.E. degree in Computer Science and Engineering in 1988 from Madurai Kamaraj University, India in the year 1988 and M.S. degree in Computer Science from BITS Pilani, India in 1993. He is currently pursuing his research at VMKV University, India .He has 21 years of Teaching experience. His areas of interest include Data structures, Multimedia Networks and Embedded Network Processor.

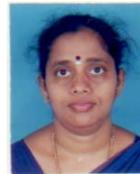

**P. Seethalakshmi** has received her B.E. degree in Electronics and Communication Engineering in 1991 and M.E. degree in Applied Electronics in 1995 from Bharathiar University, India. She obtained her doctoral degree from Anna University Chennai, India in the year 2004. She has 15 years of Teaching experience and her areas of research includes Multimedia Streaming, Wireless Networks, Network Processors and Web Services.